\documentclass[aps,pra,twocolumn,amsmath,amssymb,groupedaddress,longbibliography]{revtex4-1}
\usepackage{graphicx}
\usepackage{dcolumn}
\usepackage{CJK}
\usepackage{bm}
\usepackage[pdfstartview=FitH, CJKbookmarks=true, bookmarksnumbered=true, bookmarksopen=true, colorlinks=true, pdfborder=001, citecolor=blue, linkcolor=blue, linktocpage=true] {hyperref}

\begin{document}

\title{Quantum metrology via chaos in a driven Bose-Josephson system}

\author{Wenjie Liu$^{1,2}$}

\author{Min Zhuang$^{1,2}$}

\author{Bo Zhu$^{1}$}

\author{Jiahao Huang$^{1}$}

\altaffiliation{Email: hjiahao@mail2.sysu.edu.cn, eqjiahao@gmail.com}

\author{Chaohong Lee$^{1,2,3}$}

\altaffiliation{Email: lichaoh2@mail.sysu.edu.cn, chleecn@gmail.com}

\affiliation{$^{1}$Guangdong Provincial Key Laboratory of Quantum Metrology and Sensing $\&$ School of Physics and Astronomy, Sun Yat-Sen University (Zhuhai Campus), Zhuhai 519082, China}

\affiliation{$^{2}$State Key Laboratory of Optoelectronic Materials and Technologies, Sun Yat-Sen University (Guangzhou Campus), Guangzhou 510275, China}

\affiliation{$^{3}$Synergetic Innovation Center for Quantum Effects and Applications, Hunan Normal University, Changsha 410081, China}

\begin{abstract}
Entanglement preparation and signal accumulation are essential for quantum parameter estimation, which pose significant challenges to both theories and experiments.
Here, we propose how to utilize chaotic dynamics in a periodically driven Bose-Josephson system for achieving a high-precision measurement beyond the standard quantum limit (SQL).
Starting from an initial non-entangled state, the chaotic dynamics generates quantum entanglement and simultaneously encodes the parameter to be estimated.
By using suitable chaotic dynamics, the ultimate measurement precision of the estimated parameter can beat the SQL.
The sub-SQL measurement precision scaling can also be obtained via specific observables, such as population measurements, which can be realized with state-of-art techniques.
Our study not only provides new insights for understanding quantum chaos and quantum-classical correspondence, but also is of promising applications in entanglement-enhanced quantum metrology.

\end{abstract}

\date{\today}
\maketitle

\section{Introduction\label{Sec1}}
Quantum metrology promises high-precision measurements for various parameters with far reaching implications for science and technology~\cite{VGiovannetti2004,VGiovannetti2006,VGiovannetti2011}.
In general, the standard procedure of parameter estimation consists of three stages: initialization, parameter-dependent time-evolution, and  measurement~\cite{BMEscher2011}.
In traditional protocols, for separable initial states~\cite{JMRadcliffe1971,WMZhang1990,XWang2003}, i.e., non-entangled states, the related measurement precision scales as the SQL, which is inversely proportional to the square root of particle number $\sqrt{N}$.
While for entangled initial states, e.g., Greenberger-Horne-Zeilinger (GHZ) state~\cite{DBouwmeester1999,WDur2000,CLee2006} or NOON state~\cite{ANBoto2000,HLee2002,MZwierz2010}, the measurement precision can be improved to the well-known Heisenberg limit (HL), which is inversely proportional to $N$.

Thus, in the long-standing quest for achieving high-precision measurement, a key goal as well as a main challenge is to prepare the entangled states and make use of the entanglement to improve measurement precision.
On one hand, entangled states often require a lot of time to generate which makes it hard to prepare~\cite{ADLudlow2015,LPezze2018,JHuang20181,JHuang20182}.
On the other hand, the entangled states are extremely fragile under the environmental noises, which inevitably decreases the measurement precision~\cite{DemkowiczDobrzanski2012,Chaves2013,JHuang2015}.

Recently, to fully utilize the temporal resources, schemes on concurrent entanglement generation and interrogation are proposed~\cite{AJHayes2018,SAHaine2020}.
Different from the traditional protocols where interrogation takes place after state-preparation, the concurrent state-preparation and interrogation devote all the time for parameter-encoding, which yields better measurement precision under the same temporal resource~\cite{AJHayes2018}.
In particular, by combining the conventional one-axis twisting (OAT) dynamics with a machined-designed sequence of rotations, a higher sensitivity of estimated parameter can be achieved compared with traditional schemes~\cite{SAHaine2020}.

Classical chaos, a well-defined property in nonlinear dynamical systems, is generally characterized by exponential instability due to sensitivity to initial conditions~\cite{MTabor1989,MCGutzwiller1990,RCHilborn2000,FritzHaake2010}.
On the other hand, entanglement is a unique property of quantum systems~\cite{RHorodecki2009}.
Up to now, numerous works aim to find out the connections between chaos and entanglement~\cite{WHZurek1994,APiga2019,CNeill2016}.
It has been explored mostly in the semiclassical regime through studies of various models such as quantum kicked top~\cite{FHaake1987,RSchack1994,SGhose2008}, Bose-Josephson junction~\cite{RUtermann1994,CLeeWH2001,EBoukobza2010}, Dicke model~\cite{PWMilonni1983,KFuruya1998,CEmary2003}, Bose-Hubbard model~\cite{ARKolovskyA2004,ARKolovskyA2016,AGiraldo2020}, and so on.
These studies are beneficial not only for the fundamental understanding of quantum-classical correspondence, but also for implications in quantum metrology where entanglement is used as an important resource~\cite{Huang2014}.

The chaotic behaviors, as an entanglement-generating dynamics, have potential to enhance the measurement precision.
In a kicked top with a Dirac-delta driving, one finds chaotic behaviors can be employed to realize a high-precision parameter estimation according to the analysis of quantum Fisher information (QFI) and Fisher information (FI)~\cite{LJFiderer2018}.
The ultimate measurement precision can be substantially enhanced by nonlinearly kicking the spin during the parameter-encoding precession and driving it into a chaotic regime.
However, despite its simplicity, such a discontinuous driving is not easy to realize~\cite{PNurwantoro2019}.

Periodically continuous time-dependent modulation allows one to manipulate quantum system in a controlled way, which is more feasible in experiments.
For example, a harmonically driven Bose-Josephson system is also a suitable platform for studying quantum chaotic dynamics.
This is what motivates us to propose a chaotic quantum metrology scheme based on the harmonically driving in Bose-Josephson system.
Naturally, based on a driven Bose-Josephson system, the following questions are worthy for considering:
(i) is it possible to utilize a more experimentally feasible continuous driving for chaotic generation and use it for entanglement-enhanced quantum metrology?
(ii) can the ultimate measurement precision bound reach beyond the SQL? and
(iii) can we find suitable realistic observable measurement for parameter estimation to demonstrate our scheme?

In this paper, we study the chaotic dynamics in a harmonically driven Bose-Josephson system, and demonstrate how to make use of the chaotic behaviors to achieve a measurement precision beyond SQL.
Unlike the conventional parameter estimation schemes, we initialize the system into a spin coherent state (SCS) and let it undergo a chaotic dynamics governed by a parameter-dependent Hamiltonian.
During the time-evolution, not only an entangled final state is generated via chaotic dynamics, but also the estimated parameter is encoded into the final state, which can effectively improve the measurement precision.
We employ a mean-field approximation to arrive at classical Poincar$\acute{e}$ sections and then identify the parameter regime at which chaotic seas appear.
Classical chaos facilitates our search for its quantum counterparts, such as linear entropy, fidelity and QFI.
By appropriately choosing the initial states, the scaling analysis of QFI reveals that chaos contributes to measurement precision enhancement.
More specifically, a sub-SQL $N$-scaling can be extracted from FIs as well as half-population difference measurement.

This paper is organized as follows.
In Sec.~\ref{Sec2}, we briefly describe the harmonically driven Bose-Josephson system.
In Sec.~\ref{Sec3}, we show how to perform quantum parameter estimation in the considered system.
In Sec.~\ref{Sec4}, we derive the approximate mean-field Hamiltonian in the classical limit, and calculate the corresponding Poincar$\acute{e}$ sections to identify the locations of the chaotic seas with several different system parameters.
In Sec.~\ref{Sec5}, under the guidance of quantum counterparts, we verify that QFI, FIs and half-population difference all exhibit an $N$-scaling beyond the SQL in a wide parameter range.
In Sec.~\ref{Sec:end}, we give a brief summary and discussion.

\section{A harmonically driven Bose-Josephson system \label{Sec2}}
We are interested in a driven Bose-Josephson system whose time-dependent Hamiltonian reads
\begin{equation}\label{BJJ}
\hat{H}(t)/\hbar=\hat{H}_1+\hat{H}_2(t),
\end{equation}
where the static term
\begin{equation}\label{hsb}
\hat{H}_1=\frac{\chi}{N}\hat{S}^2_z+B_z\hat{S}_z,
\end{equation}
and the time-dependent driven term
\begin{equation}\label{hbs}
\hat{H}_2(t)=B_x\cos\omega t\hat{S}_x.
\end{equation}
Here, $\chi$ denotes the nonlinear interaction coupling strength, $B_z$ is the strength of the static longitudinal magnetic field, and $B_x\cos\omega t$ represents a harmonically driven transverse magnetic field with $B_x$ and $\omega$ the modulation amplitude and frequency, respectively.
$N$ is the total particle number of the system.
Throughout the paper, for convenience, we set $\omega=2\pi$ and $\hbar=1$.

The Bose-Josephson system can be regarded as a spin-$J$ system comprising of $N$ spin-1/2 particles  with a pseudospin length $J=N/2$.
Generally, this system can be realized by an ensemble of two-mode Bose condensed atoms~\cite{CLee2012}.
In angular momentum representation, the collective angular momentum operators $\hat{S}_{\alpha}$ with $\alpha=x,y,z$ can be expressed in terms of Pauli operators,
\begin{equation}\label{s2}
\hat{S}_{\alpha}=\sum^N_{k=1}\frac{\hat{\sigma}_{\alpha}^{(k)}}{2},
\end{equation}
which satisfy the commutation relation
\begin{equation}\label{s2}
[\hat{S}_{\alpha},\hat{S}_{\beta}]=i\hbar\epsilon_{\alpha\beta\gamma}\hat{S}_{\gamma}.
\end{equation}
The corresponding collective spin operators are defined as
\begin{eqnarray}\label{3sb}
\hat{S}_x&=\frac{1}{2}(\hat{b}^{\dag}_2\hat{b}_1+\hat{b}^{\dag}_1\hat{b}_2),\\
\hat{S}_y&=\frac{1}{2i}(\hat{b}^{\dag}_2\hat{b}_1-\hat{b}^{\dag}_1\hat{b}_2),\\
\hat{S}_z&=\frac{1}{2}(\hat{b}^{\dag}_2\hat{b}_2-\hat{b}^{\dag}_1\hat{b}_1),
\end{eqnarray}
where $\hat{b}_\mu(\hat{b}^{\dag}_\mu)$ is atom annihilation (creation) operator and
$\hat{n}_\mu=\hat{b}^{\dag}_\mu\hat{b}_\mu$ is the occupation operator in mode $\mu$.
Naturally, one can call $\hat{S}_z$ as the half-population difference operator.
$\{|J,m_z\rangle\}$ called as Dicke basis, are common eigenstates of $\hat{S}^2$ and $\hat{S}_z$ with eigenvalues $J(J+1)$ and $m_z$.
In the Dicke basis $\{|J,m_z\rangle\}$ with $m_z=-J,-J+1,...,J$, arbitrary quantum states of the above Hamiltonian~\eqref{BJJ} can be written as $|\Psi\rangle=\sum_{m_z}C_{m_z}|J,m_z\rangle$, where $C_{m_z}$ denotes the probability amplitude projecting onto the basis $|J,m_z\rangle$.

The Hamiltonian~\eqref{BJJ} is equivalent to the Hamiltonian of a two-mode Bose-Hubbard model
\begin{equation}\label{BHM}
\begin{aligned}
\hat{H}_{BH}=&\frac{\chi}{4 N}\left(\hat{n}_{2}-\hat{n}_{1}\right)^{2}+\frac{B_{z}}{2}\left(\hat{n}_{2}-\hat{n}_{1}\right) \, \\
&+\frac{B_{x} \cos \omega t}{2}\left(\hat{b}_{2}^{\dag} \hat{b}_{1}+\hat{b}_{1}^{\dag} \hat{b}_{2}\right)
\end{aligned}
\end{equation}
with $N=\langle\hat{N}\rangle=\left\langle\hat{n}_{1}+\hat{n}_{2}\right\rangle$ representing the atom number $N$.
Since $\left[\hat{N},\hat{H}_{BH}\right]=0$, the total number of atoms $N$ is conserved.
Experimentally, the Bose-Josephson system~\eqref{BJJ} can be realized via a two-mode Bose-Hubbard system consisting of trapping bosons in a double-well potential~\cite{YShin2004,TSchumm2005,MAlbiez2005}.

When $B_z=0$, the properties of ground states depending on the above Hamiltonian~\eqref{BJJ} are related to the interplay between transverse magnetic field $B_x$ and the nonlinearity strength $\chi$.
Under the condition $\omega=0$, if system is dominated by the transverse magnetic field $B_x$, the ground state is an SU(2) SCS.
For a sufficient and positive nonlinearity strength $\chi\gg0$, the ground state turns to $|J,0\rangle$ for even $N$ or $\frac{1}{\sqrt{2}}(|J,1/2\rangle+|J,-1/2\rangle)$ for odd $N$.
While if the negative nonlinearity strength dominates $\chi\ll0$, two degenerate ground states $|J,J\rangle$ and $|J,-J\rangle$ appear and any superposition of these two states is a ground state, including the GHZ state $\frac{1}{\sqrt{2}}(|J,J\rangle+|J,-J\rangle)$.

For $\omega\neq0$, the periodically driven system~\eqref{BJJ} respects a discrete translation symmetry in time domain, $\hat{H}(t+T)=\hat{H}(t)$ with a Floquet period $T=2\pi/\omega$.
$\hat{\mathcal{T}}$ is the time-ordering operator, and the time-evolution operator in a single period can be calculated as
\begin{equation}\label{ute}
\hat{U}(T;0)=\hat{\mathcal{T}}\exp\left(-i\int^T_0\hat{H}(t)dt\right)\equiv\exp(-i\hat{H}_FT).
\end{equation}
The system stroboscopic dynamics at moments $nT$ ($n=1,2,...$) is governed by the time-averaged Hamiltonian
\begin{equation}\label{staticsystem}
\hat{H}_F=\frac{i}{T}\log\hat{U}(T;0).
\end{equation}
So far, we map the time-dependent system~\eqref{BJJ} to a time-independent one~\eqref{staticsystem}, and we can obtain the stroboscopic time evolution from the static Hamiltonian $\hat{H}_F$.
The periodically modulated transverse magnetic field flexibly controls the system behaviors.
The interplay between the nonlinearity strength and modulation amplitude may generate the chaotic behaviors for a fixed modulation frequency, which will be discussed later.

\section{Quantum parameter estimation}\label{Sec3}
The longitudinal magnetic field causes a precession about the $z$-axis.
The strength of the longitudinal magnetic field $B_z$ is the parameter we want to estimate.
Assuming other parameters are known, we are interested in the measurement precision of $B_z$.
Generally, quantum parameter estimation can be described by a Positive-Operator Valued Measures (POVM) comprised of a set of positive Hermitian operators ${\hat{E}(x_n)}$, which satisfy $\hat{E}(x_n)\geq0$ and $\sum_{n}\hat{E}(x_n)=1$, and $x_n$ denotes $n$-th measurement result of the measured observable $\hat{E}(x_n)$.
Starting from an initial non-entangled state $|\psi_0\rangle$, the system can be driven to undergo chaotic dynamics while the estimated parameter $B_z$ is encoded into the evolved state.
The entangled output state satisfies $\left|\psi_{f}\left(B_{z}\right)\right\rangle=\hat{U}\left|\psi_{0}\right\rangle$
with the time-evolution operator
\begin{equation}
\hat{U}=\hat{\mathcal{T}} \exp \left[-i \int_{t_{0}}^{t} \hat{H}\left(t^{\prime}\right) d t^{\prime}\right]
\end{equation}
governed by the whole system~\eqref{BJJ}.
Here, $\hat{H}_1$ acts as an OAT Hamiltonian which can be exploited to prepare the squeezed states as well as encode the estimated parameter $B_z$.
At the same time, $\hat{H}_2$ rotates the system around the $x$-axis with a continuous driving.
Both terms in the stroboscopic time evolution conspire to provide a high-precision measurement from an initial non-entangled state.

By reading out the final state, the conditional probability of the measurement result $x_n$ given by estimated parameter $B_z$ yields
\begin{equation}
P(x_n|B_z)=Tr\left(\hat{E}(x_n) |\psi_f (B_z)\rangle\langle\psi_f (B_z)|\right).
\end{equation}
The FI related to the conditional probability reads as
\begin{equation}\label{FI}
F_I\left(B_{z}\right)=\sum_{n} P\left(x_n|B_{z}\right)\left(\frac{\partial \ln \left[P\left(x_n|B_{z}\right)\right]}{\partial B_{z}}\right)^{2}.
\end{equation}
For a given POVM, the Cram$\acute{e}$r-Rao bound is expressed as
\begin{equation}
\Delta B_z\geq1/\sqrt{\mathcal{N} F_I},
\end{equation}
which gives the minimal achievable uncertainty of the estimated parameter with $\mathcal{N}$ times trials.
FI represents an ability to measure the parameter and instructs one to enhance the measurement precision by maximizing it.
Maximizing the FI by trying all possible POVM, it is known that, the ultimate measurement precision is limited by the quantum Cram$\acute{e}$r-Rao bound
\begin{equation}\label{QCRB}
\Delta B_z\geq1/\sqrt{\mathcal{N} F_Q}.
\end{equation}
The corresponding FI with the optimal POVM measurement is called as QFI $F_Q$.
The QFI can be calculated as
\begin{equation}
F_{Q}\left(B_{z}\right)=4\left[\left\langle\psi_{f}^{\prime} | \psi_{f}^{\prime}\right\rangle-\left|\left\langle\psi_{f}^{\prime} | \psi_{f}\right\rangle\right|^{2}\right]
\end{equation}
where $\left|\psi_{f}^{\prime}\right\rangle=\frac{\partial\left|\psi_{f}\right\rangle}{\partial B_{z}}$
denotes the derivative of final state with respect to the estimated parameter $B_z$.
Obviously, the QFI is only depends on the final state as well as its derivative.

\section{Mean-field approximation}\label{Sec4}
The initial SCS reads in the Dicke basis as
\begin{equation}
\begin{aligned}
|J, \theta, \phi\rangle=&\sum_{m_z=-J}^{J} \sqrt{\left(\begin{array}{c}{2 J} \\ {J-m_z}\end{array}\right)}\sin (\theta / 2)^{J-m_z} \, \\
& \times\cos (\theta / 2)^{J+m_z} e^{i(J-m_z) \phi}|J, m_z\rangle
\end{aligned}
\end{equation}
with $\theta$ and $\phi$ as the polar and azimuthal angles respectively.
First, we should know the conditions for generating chaos and then we try to perform the measurement by using the chaotic behaviors.
To study the properties of chaos, we employ the mean-field theory to obtain a classical Hamiltonian equation.
We then examine its Poincar$\acute{e}$ sections, especially for identifying the parameter regime at which chaotic behaviors appear.
We consider Hamiltonian~\eqref{BHM}. In the semiclassical limit $N\rightarrow\infty$, the whole system is dominated by the condensed atoms and can be approached via the mean-field approximation, i.e., $\hat{b}_{\mu} \approx \psi_{\mu}, \hat{b}_{\mu}^{\dag} \approx \psi_{\mu}^{*}$,
with $\psi_{\mu}=\left\langle\hat{b}_{\mu}\right\rangle, \psi_{\mu}^{*}=\left\langle\hat{b}_{\mu}^{\dag}\right\rangle$.
The total particle number $\left|\psi_{1}\right|^{2}+\left|\psi_{2}\right|^{2}=N$ is conserved.
Applying the mean-field approximation, the classical Hamiltonian equation is written as
\begin{equation}
\begin{aligned}
H_{M F}=&\frac{\chi}{4 N}\left(n_{2}-n_{1}\right)^{2}+\frac{B_{z}}{2}\left(n_{2}-n_{1}\right) \, \\
&+\frac{B_{x} \cos \omega t}{2}\left(\psi_{1}^{*} \psi_{2}+\psi_{2}^{*} \psi_{1}\right).
\end{aligned}
\end{equation}
Substituting the mean-field Hamiltonian into the equation
\begin{equation}
i \hbar \frac{d \psi_{\mu}(t)}{d t}=\frac{\partial H_{M F}}{\partial \psi_{\mu}^{*}(t)},
\end{equation}
the two coupled differential equations yield
\begin{equation}
\begin{aligned}
i \frac{d \psi_{1}}{d t}=&-\frac{B_{z}}{2} \psi_{1}+\frac{ \chi}{2 N}\left(\left|\psi_{1}\right|^{2}-\left|\psi_{2}\right|^{2}\right) \psi_{1}+\frac{B_{x} \cos \omega t}{2} \psi_{2}, \, \\
i \frac{d \psi_{2}}{d t}=&+\frac{B_{z}}{2} \psi_{2}+\frac{ \chi}{2 N}\left(\left|\psi_{2}\right|^{2}-\left|\psi_{1}\right|^{2}\right) \psi_{2}+\frac{B_{x} \cos \omega t}{2} \psi_{1}.
\end{aligned}
\end{equation}
The complex amplitudes can be expressed as $\psi_{\mu}=\sqrt{n_{\mu}} \exp \left(i \phi_{\mu}\right)$ in terms of the particle number $n_\mu=\psi_{\mu}^{*}\psi_{\mu}$ and the phase $\phi_\mu$.
Due to two degrees of freedom lied in system,
one can further set up two canonical variables, the fractional population imbalance $z=\frac{n_{1}-n_{2}}{N}=\cos\theta$ and the relative phase $\phi=\phi_1-\phi_2$.
The equations of motion in parameter space $(\phi,z)$ are given as
\begin{equation}\label{motionequation}
\begin{array}{l}\displaystyle{\frac{d \phi}{d t}=z \chi-\frac{z B_{x} \cos \omega t}{\sqrt{1-z^{2}}} \cos \phi-B_{z}}, \\
{\frac{d z}{d t}=B_{x} \cos \omega t \sqrt{1-z^{2}} \sin \phi}.
\end{array}
\end{equation}
We numerically obtain the Poincar$\acute{e}$ sections by recording the phase space locations at each integer multiple of period $T$.

\begin{figure}[htp]
\begin{center}
\includegraphics[width=0.45\textwidth]{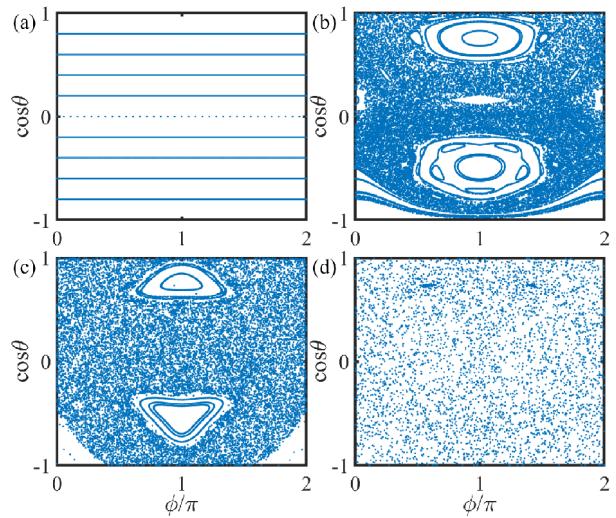}
\end{center}
\caption{Poincar$\acute{e}$ sections governed by Eq.~\eqref{motionequation} at different strengths of the transverse magnetic field: (a) $B_x=0$, (b) $B_x=1.5$, (c) $B_x=3$ and (d) $B_x=5.5$. The other parameters are chosen as $N=1000$, $\chi=10$ and $B_z=\pi/2$.
}
\label{fig:1}
\end{figure}

The mean-field approach enables us to locate parameter regimes at which chaotic and regular regions coexist or chaotic behaviors dominate. Once such parameter regimes are identified, we then explore the difference between initial states locating in different regions.
Through plotting the Poincar$\acute{e}$ sections, we get the trajectory of the system up to $500$ periods for different transverse magnetic field strengths $B_x$: $0$, $1.5$, $3$ and $5.5$, as depicted in Fig.~\ref{fig:1}.
Here, the particle number is $N=1000$, the longitudinal magnetic field strength is $B_z=\pi/2$ and the nonlinearity strength is $\chi=10$.

In the absence of the transverse magnetic field, the regular behaviors perfectly emerge in the whole phase space, see Fig.~\ref{fig:1}~(a).
Once the transverse magnetic field is added, in addition to the regular behaviors, the chaotic behaviors also appear, see Fig.~\ref{fig:1}~(b).
When $B_x$ increases further, the chaotic regions enlarge while the regular regions shrink, see Fig.~\ref{fig:1}~(c).
Until to a sufficiently large $B_x=5.5$, the chaotic regions nearly dominate, see Fig.~\ref{fig:1}~(d).
In addition, we note that these Poincar$\acute{e}$ sections are almost symmetric about the line $\phi=\pi$, and it allows us to just concentrate on its left part.

\section{Full quantum approach}\label{Sec5}
The chaotic behaviors are closely related to the entanglement generation.
We devote to seek suitable quantum counterparts to find the connections to the corresponding chaos.
Then, we also calculate the corresponding QFI, which can be used as a measure for the precision bounds of estimated parameter.
We find that, after a long-time evolution, the QFI for the chaotic region will be much larger than the one for the regular region.
Furthermore, by choosing initial coherent states in different regions, we evaluate the scalings of QFI versus evolution time $t$ and total particle number $N$.
The scaling of QFI with chaotic dynamics can exceed the SQL.
To demonstrate our scheme from the perspective of experimental realization, for the case of mostly chaotic phase space, we also analyze the scalings of practical observables, such as FI and half-population difference.

\subsection{In mixed phase space}\label{subsec51}

\begin{figure}[htp]
\begin{center}
\includegraphics[width=0.45\textwidth]{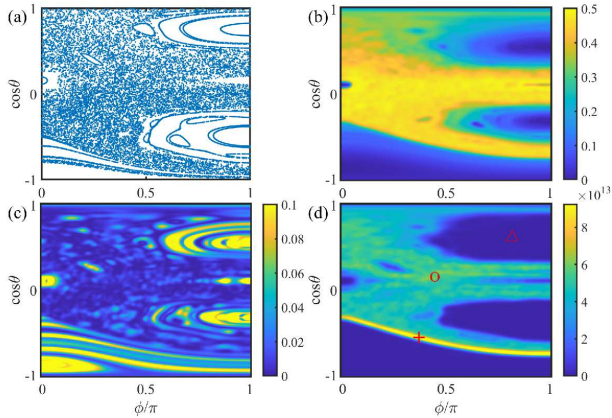}
\end{center}
\caption{(a) Poincar$\acute{e}$ section as a function of $\theta,\phi$ corresponding to Fig.~\ref{fig:1}~(b).
(b)-(d) Its corresponding phase space distributions of linear entropy, fidelity and QFI in the quantum setting exemplarily for system with large particle number $(N=1000)$ up to $t=2^{15}T$.
The other parameters are chosen as $\chi=10$, $B_x=1.5$ and $B_z=\pi/2$.
Marks labeled in (d) are chosen as initial parameters in Fig.~\ref{fig:3}.
}
\label{fig:2}
\end{figure}

First, we study the quantum counterparts of classical phase space in which chaotic regions coexist with regular regions and analyze the scalings of QFI.
Due to the symmetry of Poincar$\acute{e}$ section about the line $\phi=\pi$, it is reasonable to just take into account the case for $\phi$ ranging from $0$ to $\pi$, as shown in Fig.~\ref{fig:2}~(a).
In order to explore the whole evolution of quantum counterparts,
all points in the Poincar$\acute{e}$ section are taken as initial states to evolve.
Numerically, we deal with the time-evolution operator $\hat{U}(T;0)$ by a series of discrete time steps $\delta t=T/1000$, and then let system evolves in time domain with a period-averaged propagator.

Entropy, as a pure quantum resource, is a powerful bridge between classical and quantum worlds.
The linear entropy characterizes the entanglement between a single particle and the rest of the system defined as
\begin{equation}
S(n T)=\frac{1}{2}\left(1-\frac{\left\langle \hat{S}_{x}\right\rangle^{2}+\left\langle \hat{S}_{y}\right\rangle^{2}+\left\langle \hat{S}_{z}\right\rangle^{2}}{J^{2}}\right)
\end{equation}
at the stroboscopic time $nT$ $(n=1,2,...,N)$.
The linear entropy nicely reproduces a similar structure to the Poincar$\acute{e}$ section, in which high entropy in chaotic region while low one in regular region, see Fig.~\ref{fig:2}~(b).
From the analysis of the linear entropy, we know the entanglement tends to increase by means of chaos.

\begin{figure}[htp]
\begin{center}
\includegraphics[width=0.45\textwidth]{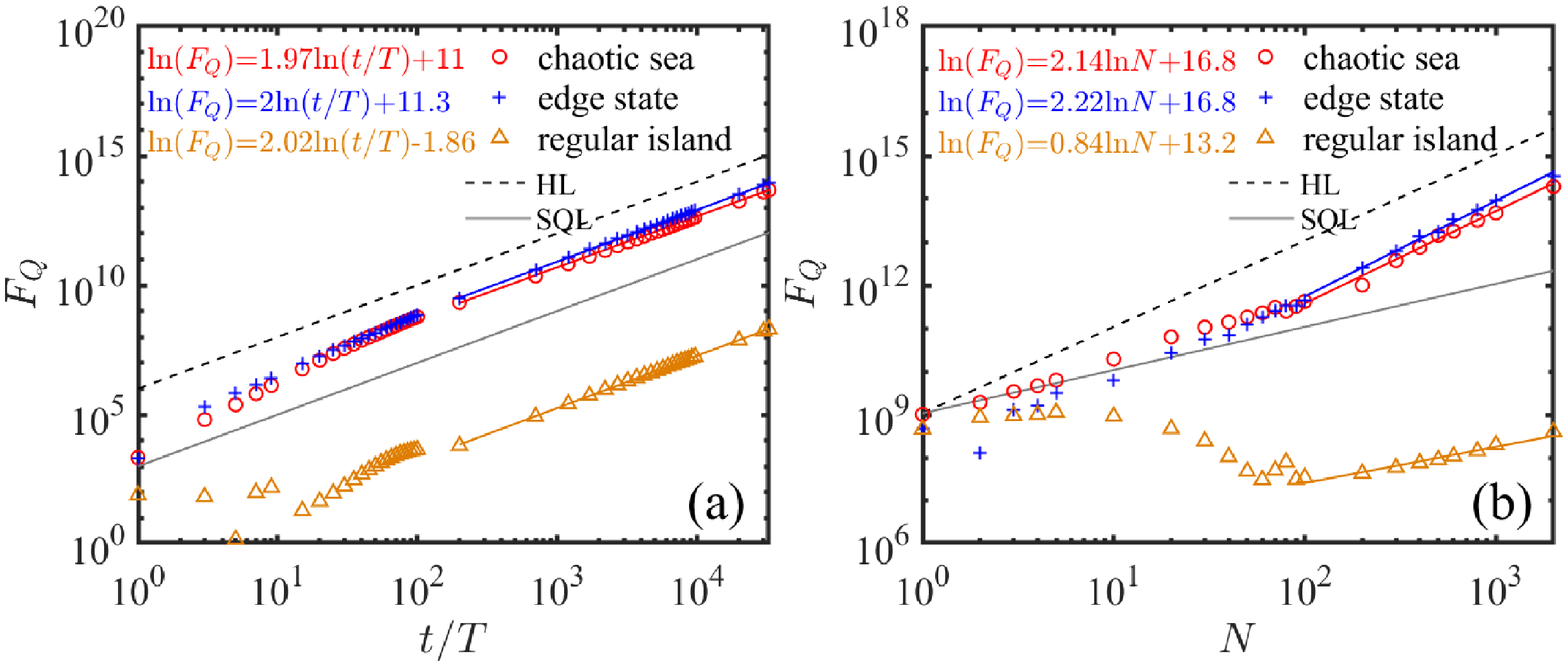}
\end{center}
\caption{QFI with respect to $t$ and $N$ for different initial parameters.
(a) The stroboscopic evolution of QFI up to $2^{15}$ periods with a system size $N=1000$.
Fits exhibit a slope of 1.97 (chaotic sea), 2 (edge state) and 2.02 (regular island) for $t$-scaling.
(b) QFI for different $N$ with a fixed evolution time $t=2^{15}T$.
Fits exhibit a slope of 2.14 (chaotic sea), 2.22 (edge state) and 0.84 (regular island) for $N$-scaling.
The other parameters are chosen as $\chi=10$, $B_x=1.5$ and $B_z=\pi/2$.
Gray line and black dashed line represent the SQL and HL, respectively.}
\label{fig:3}
\end{figure}

The fidelity is given as
\begin{equation}
F(nT)=\langle\psi(0) | \psi(n T)\rangle|^{2},
\end{equation}
which is used to quantify how much information remains in the time-evolved state $|\psi(nT)\rangle$ comparing to the initial state $|\psi(0)\rangle$.
A good quantum-classical correspondence is exhibited via fidelity, see Fig.~\ref{fig:2}~(c).
Initial SCS located on chaotic region tends to undergo an ergodic dynamics, thus losing a lot of local information in final state with a low fidelity.
Whereas for the initial state in regular region, the system remains a relatively high fidelity after a periodic dynamics.
The linear entropy and fidelity not only both establish a good correspondence between the quantum dynamics and classical phase space, they also present a complementary relationship, that is, high entropy matches low fidelity and low entropy corresponds to high fidelity.

The generated entanglement can be exploited for metrology, which explains how chaos contributes to the enhancement of measurement precision.
More directly, we calculate the corresponding QFI, which is closely related to quantum parameter estimation according to Eq.~\eqref{QCRB}.
The similar structure reflects that the classical dynamics of the system has an excellent correspondence with the quantum dynamics of QFI, in Fig.~\ref{fig:2}~(d).
After a long-time evolution, final QFIs arising from initial states in chaotic or regular regions are obviously different.
Chaotic sea takes on a large QFI while the regular island possesses a relatively small QFI.
Remarkably, the boundary between the chaotic sea and regular island (called as edge state) has a larger QFI.
We name these three typical regions as chaotic sea (circle), edge state (cross) and regular island (triangle), which are marked in Fig.~\ref{fig:2}~(d).

The distributions in Bloch sphere of final states reveal an ergodic dynamics for chaotic sea or edge state and a localized dynamics for regular island (see Appendix~\ref{appendixA} for more details).
It is shown that, the linear entropy, fidelity and QFI can all be used as counterparts for quantum-classical correspondence.
In comparison with the classical Poincar$\acute{e}$ section, these quantum counterparts can not only distinguish regular and chaotic regions, but also present a much richer results by quantifying the chaos from different perspectives.
The particle number has influence on the value of these three quantum counterparts, and a scaling analysis is exhibited in
Appendix~\ref{appendixB}.

In order to investigate the role of initial SCS in parameter estimation, still within the mixed phase space, we adopt the initial parameters in different regions in Fig.~\ref{fig:2}~(d) to explore their differences, from the perspective of QFI.
In Fig.~\ref{fig:3}, different marks correspond to the different states initially prepared in the locations shown in Fig.~\ref{fig:2}~(d).
Fig.~\ref{fig:3}~(a) shows, beginning from a moderate evolved time, QFI does still increase with $t$ and shows a nearly quadratic $t$-scaling for chaotic, edge and regular initial situations.
When initial state locates on chaotic sea (red circle) or edge state (blue cross), QFI evolves with time between SQL and HL, while for the initial state locating within a regular island (yellow triangle), QFI evolves below the SQL.
In addition, we note that edge state is superior to chaotic sea all the time.
The evolved time as a resource is able to improve its value of QFI.
For a fixed evolution time $t=2^{15}T$, Fig.~\ref{fig:3}~(b) reflects that the $N$-scaling of QFI is sensitive to the initial states.

Besides the early decay in regular region, QFIs increase with the particle number $N$ for chaotic, edge and regular situations.
The numerical results reveal that initial state in the chaotic sea performs best for small $N$, while for larger system size ($N>100$) edge state performs best.
Fits for three different initial states take on a slope of 2.14 (chaotic sea), 2.22 (edge state) and 0.84 (regular island) for the $N$-scaling.
Both $N$-scalings for edge state and chaotic sea can approach even better than HL.
While for the initial state in regular island fails to beat the SQL.
It is indicated that, chaos as a kind of resource, allows one to attain a high-precision measurement beyond SQL.

\subsection{In fully chaotic regime}\label{subsec52}

According to the scaling analysis in the mixed phase space, we know that chaotic behaviors can play an important role for quantum parameter estimation.
We naturally turn to the fully chaotic regime, and evaluate the scalings for QFI with respect to $t$ and $N$.
In particular, as the system enters a full chaos, the information about initial SCS is rapidly lost.
Therefore the initial SCS can be chosen in anywhere without changing much the QFI, the situation is markedly different from the case of a mixed phase space.

\begin{figure}[htp]
\begin{center}
\includegraphics[width=0.45\textwidth]{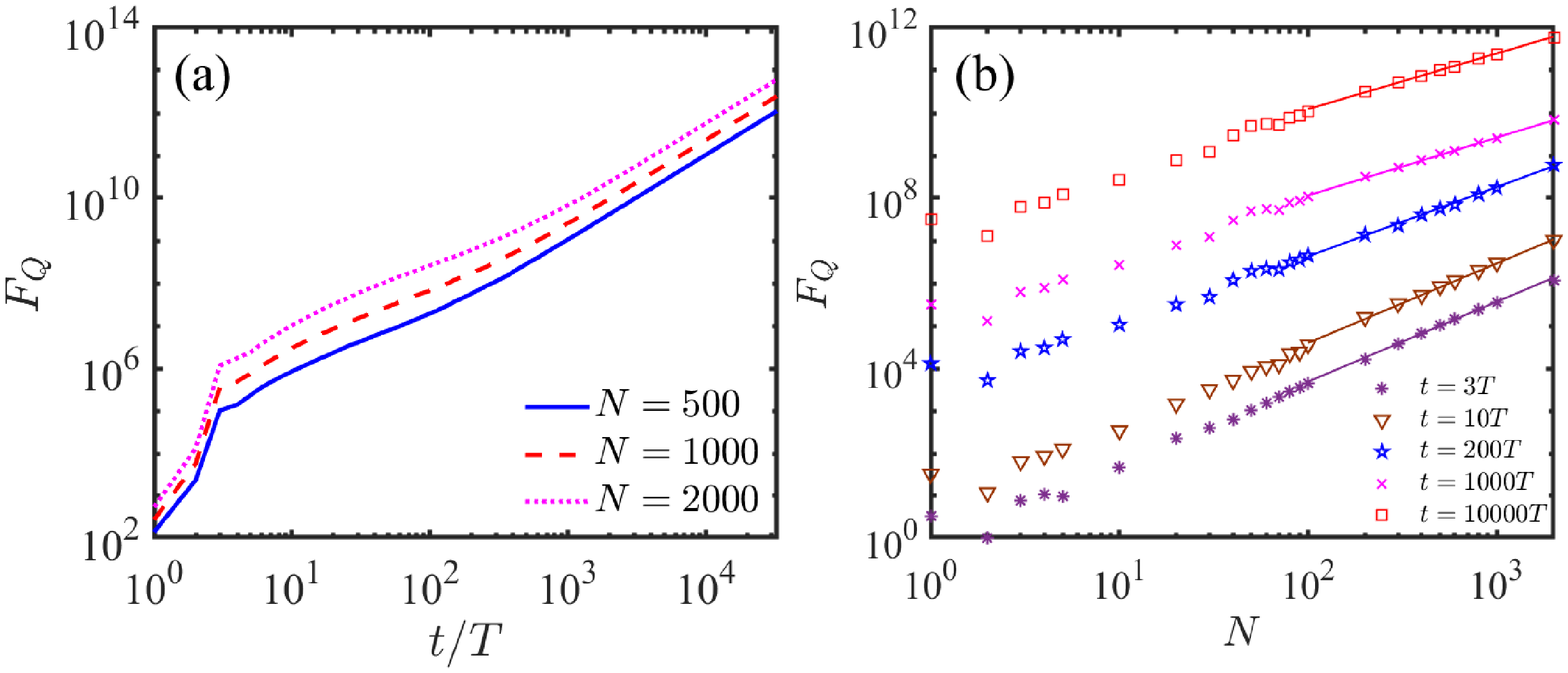}
\end{center}
\caption{(a) QFI as a function of stroboscopic evolution time with different system sizes $N$=500 (solid line), 1000 (dashed line) and 2000 (dotted line).
(b) QFI as a function of system sizes for different evolution time.
Fits exhibit slopes 1.88 ($t=3T$), 1.87 ($t=10T$), 1.63 ($t=200T$), 1.35 ($t=1000T$) and 1.31 ($t=10000T$) for $N$-scaling.
The initial state locates on $\theta=2.423,\phi=1.126$.
The other parameters are chosen as $\chi=10$, $B_x=5.5$ and $B_z=\pi/2$.}
\label{fig:4}
\end{figure}

Below we focus on phase space of QFI in which its classical counterpart displays fully chaotic behaviors corresponding to Fig.~\ref{fig:1}~(d).
Independent of the details of the initial state, we randomly choose the initial parameters $\theta=2.423,\phi=1.126$.
In the fully chaotic case ($\chi=10$, $B_x=5.5$ and $B_z=\pi/2$), the numerical results in Fig.~\ref{fig:4}~(a) depict a time evolution of QFI.
QFI drastically increases during a short evolution time ($t\simeq3T$) and behaves similarly at three different system sizes: $N=500$, 1000, 2000.
For a large $N$, QFI is more sensitive to the early time.
In order to further explore the influence of the evolution time, we consider system size $N=1000$ and plot QFI as the system size $N$ is varied in Fig.~\ref{fig:4}~(b) for different evolution times: $t=3T$, $10T$, $200T$, $1000T$ and $10000T$.
Other parameters agree with the Poincar$\acute{e}$ section shown in Fig.~\ref{fig:1}~(d).
Fig.~\ref{fig:4}~(b) reveals the $N$-scaling for QFI nearly achieves HL for short evolution time ($t\simeq3T$), then gradually decays to SQL with increasing evolution time.

\begin{figure}[htp]
\begin{center}
\includegraphics[width=0.45\textwidth]{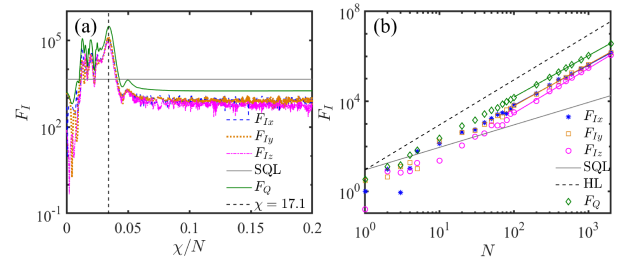}
\end{center}
\caption{(a) QFI and FIs of collective spin operator $\hat{S}_x$, $\hat{S}_y$ and $\hat{S}_z$ for various nonlinearity strengths $\chi$.
The system size is chosen as $N=500$.
(b) QFI and FIs as a function of $N$ for a fixed $\chi=17.1$.
Fits exhibit slopes 1.84 ($F_Q$), 1.79 ($F_{Ix}$), 1.84 ($F_{Iy}$) and 1.99 ($F_{Iz}$) for $N$-scaling.
The initial state locates on $\theta=2.423,\phi=1.126$ and evolution time is $t=3T$.
The other parameters are chosen as $B_x=5.5$ and $B_z=\pi/2$.
Gray line and black dashed line represent the SQL and HL, respectively.}
\label{fig:5}
\end{figure}

In principle, QFI corresponds to the optimal POVM measurement and just mathematically sets the ultimate bound of the measurement precision, but it may not always be realistic.
To approach the precision bounds set by QFI, we specify a feasible measurement, i.e., FI~\eqref{FI}.
Starting from a given initial state $|J,\theta,\phi\rangle$, where $\theta$ and $\phi$ signify the position of the SCS, the estimated parameter $B_z$ is encoded into the evolved state during a stroboscopic evolution.
The Dicke basis $\{|J,m_{\alpha}\rangle\}$ comprises of the common eigenstates of $\hat{S}^2$ and $\hat{S}_{\alpha}$ for $m_{\alpha}=-J,-J+1,...,J$ with $\alpha=x,y,z$.
The obtained final state can be written in the Dicke basis as $|\psi_f\rangle=\sum^J_{m_{\alpha}=-J}C_{m_{\alpha}}(B_z)|J,m_{\alpha}\rangle$.
The conditional probabilities related to the collective spin operator $\hat{S}_{\alpha}$ can be calculated as
$P\left(m_{\alpha} | B_{z}\right)=\left|C_{m_{\alpha}}\left(B_{z}\right)\right|^{2}$, and its FI is defined as
\begin{equation}\label{FIZ}
F_{I\alpha}\left(B_{z}\right)=\sum_{m_{\alpha}} \frac{1}{\left|C_{m_{\alpha}}\left(B_{z}\right)\right|^{2}}\left(\frac{\partial\left|C_{m_{\alpha}}\left(B_{z}\right)\right|^{2}}{\partial B_{z}}\right)^{2}.
\end{equation}
\begin{figure}[htp]
\begin{center}
\includegraphics[width=0.45\textwidth]{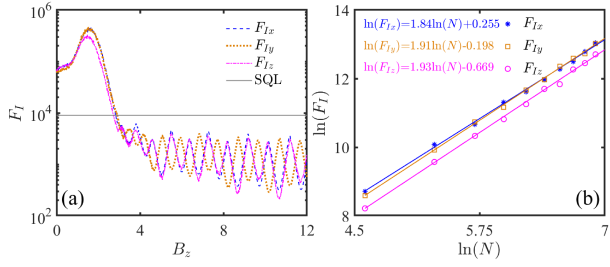}
\end{center}
\caption{(a) FIs as a function of estimated parameter.
The system size is  $N=1000$.
(b) A ln($F_I$)-ln($N$) scaling analysis for the optimal $B_z$ corresponding its maximum $F_I$ in (a).
The initial state locates on $\theta=2.423,\phi=1.126$ and evolution time is $t=3T$.
The other parameters are chosen as $\chi=17.1$ and $B_x=5.5$.
Gray line represents the SQL.}
\label{fig:6}
\end{figure}

Due to the important role of nonlinearity strength $\chi$ in chaotic behaviors, we also devote to calculate the QFI and FI versus $\chi$ to find out the suitable $\chi$.
Based on the aforementioned experience in Fig.~\ref{fig:4}, we consider the system size $N=500$ and perform time evolution up to $t=3T$.
Moreover, other parameters are the same as those in Fig.~\ref{fig:4}.
By calculating the FIs related to collective spin operators $\hat{S}_x$, $\hat{S}_y$ and $\hat{S}_z$, we attain a sensitivity over SQL in a certain region of nonlinearity strength $\chi$ but always below its QFI, see Fig.~\ref{fig:5}~(a).
It allows us to identify the optimal nonlinearity strength locating in the vicinity of $\chi=17.1$.
Then, we hope to verify that the obtained parameter regime indeed respects the anticipated scaling beating the SQL.
By further setting $\chi=17.1$, we numerically obtain the scaling of QFI and FIs with respect to $N$.
The $N$-scaling regarding QFI and FIs obviously beyond the SQL and near HL shown in Fig.~\ref{fig:5}~(b).

Depending on known results in Fig.~\ref{fig:5}, we further analyze the optimal estimated parameter $B_z$ as well as its achievable scaling at a fixed nonlinearity strength $\chi=17.1$.
We consider $N=1000$ and other parameters the same as those in Fig.~\ref{fig:5}.
In Fig.~\ref{fig:6}~(a), one can see that FIs with increasing $B_z$ all exhibit a similar behavior as it gradually rises to a maximum value, then gradually decays under the SQL, and finally tends to a small oscillation.
FIs outperform the SQL in a broad range ($0\leq B_z\leq2.8$), which includes the above discussed value $B_z=\pi/2$.
Furthermore, we perform a scaling analysis to confirm that the obtained estimated parameter regime indeed shows the anticipated scaling to improve the measurement precision.
The system size spans from $N=100$ to $N=1000$, and we choose the optimal $B_z$ corresponding to the maximum $F_I$ in Fig.~\ref{fig:6}~(a).
These fits yield a slope of 1.84 ($F_{Ix}$), 1.91 ($F_{Iy}$), and 1.93 ($F_{Iz}$) for $N$-scaling, all nearly reach the HL, see Fig.~\ref{fig:6}~(b).

\begin{figure}[htp]
\begin{center}
\includegraphics[width=0.45\textwidth]{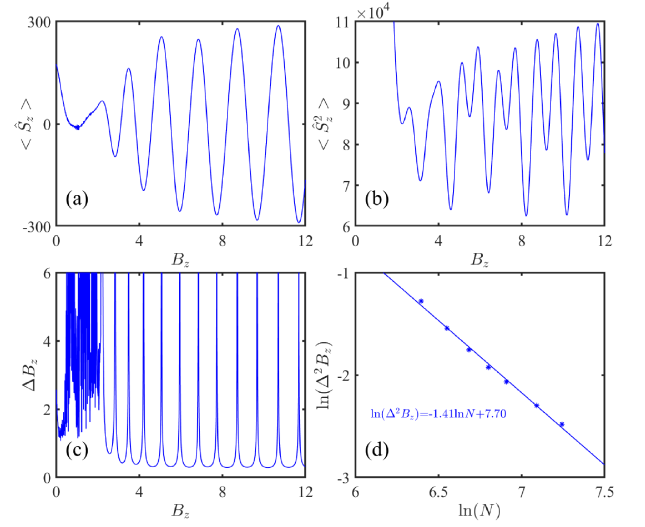}
\end{center}
\caption{(a)-(c) The expectative value of half-population difference operator as well as its square and the corresponding parameter uncertainty for different estimated parameters $B_z$.
The system size is chosen as $N=1400$.
(d) A ln($\Delta^2 B_z$)-ln($N$) scaling analysis for the optimal estimated parameter $B_z$ corresponding the minimum $\Delta {B_z}$ in (c).
The system size takes value from $N=600$ to 1400.
The initial state locates on $\theta=2.423$, $\phi=1.126$ and evolution time is $t=3T$.
The other parameters are chosen as $\chi=17.1$ and $B_x=5.5$.
}
\label{fig:7}
\end{figure}

Finally, we proceed to carry out quantum precision calculations by half-population difference measurement.
According to the error propagation formula,
\begin{equation}
\Delta B_{z}=\frac{(\Delta \hat{O})_{f}}{\left|\partial\langle\hat{O}\rangle_{f} / \partial B_{z}\right|},
\end{equation}
where $\hat{O}$ and $\langle\hat{O}\rangle_f=\langle\psi_f|\hat{O}|\psi_f\rangle$ respectively denote an arbitrary observable as well as its expectation value onto the final state $|\psi_f\rangle$.
The corresponding standard deviation is given as $(\Delta \hat{O})_{f}=\sqrt{\left\langle\psi_{f}\left|\hat{O}^{2}\right| \psi_{f}\right\rangle-\left\langle\psi_{f}|\hat{O}| \psi_{f}\right\rangle^{2}}$.

We choose the half-population difference $\hat{S}_z$ as the observable, which is easy to implement in experiments.
Figs.~\ref{fig:7}~(a) and (b) manifest the estimated parameter dependence of $\langle\hat{S}_{z}\rangle$ and $\langle\hat{S}^2_{z}\rangle$ for $N=1400$, $t=3T$, $\chi=17.1$ and $B_x=5.5$.
Note that, the initial state is still stabilized at $\theta=2.423$, $\phi=1.126$.
A sharp changing of observable $\hat{S}_z$ about parameter $B_z$ contributes to the parameter precision.
Unlike the irregular form when $B_z$ lies in small values, the expectative values of $\langle\hat{S}_{z}\rangle$ and $\langle\hat{S}^2_{z}\rangle$  present relatively regular oscillations until to a moderate strength, which has potential to effectively estimate $B_z$.

The standard deviation of the estimated parameter associated by $\hat{S}_z$ shown in Fig.~\ref{fig:7}~(c) yields a a mess at the beginning, while a regular form arises as the increasing of $B_z$.
After extracting the optimal $B_z$ for a minimum $\Delta B_z$ in Fig.~\ref{fig:7}~(c), we evaluate the corresponding uncertainty scaling $\Delta^2 B_z$ versus $N$.
Taking value of particle number from $N=600$ to $1400$, a ln($\Delta^2 B_z$)-ln($N$) scaling analysis is shown in Fig.~\ref{fig:7}~(d), where the slope is found to be -1.41, beating the SQL.

We provide numerical evidences that ultimate achievable precision strictly outperforms the optimal classical strategy via the chaotic dynamics scheme.
Our scheme can be demonstrated in experiments.
Not only a high-precision parameter estimation can be realized via the FIs, but also a $N$-scaling beyond SQL appears via the half-population difference measurement.

\section{Summary and Discussion}\label{Sec:end}

Based on a driven Bose-Josephson system, we propose a dynamic high-precision measurement scheme, which generates quantum entanglement via chaos and simultaneously encodes the parameter to be estimated.
Our scheme not only overcomes the challenges of entangled state preparation, but also utilizes the most of temporal resources.
For the first time to our knowledge, we build a completely quantum-classical correspondence between the classical Poincar$\acute{e}$ section and three observables: the linear entropy, the fidelity and the QFI.
Remarkably, these three quantum counterparts well reflect the coexistence of chaotic and regular dynamics from different perspectives.
Further, we respectively extract three typical initial states in the mixed phase space in which chaotic behavior and regular one coexist to explore the scaling properties.
The scaling analysis in mixed phase space instructs us to enhance the parameter estimation precision via the chaotic dynamics.
Then we turn to a fully chaotic case to investigate the corresponding QFI and FIs satisfying scalings beyond the SQL.
Not limited to the QFI and FIs, we also harvest a scaling beating the SQL via a feasible observable, half-population difference measurement.

We have demonstrated that chaotic dynamics contributes to the precision of estimated parameter in a closed system.
However, decoherence, such as dephasing and dissipation, is an important ingredient in realistic systems and it is meaningful to extend our scheme to open systems.
In practice, many physical systems face the simultaneous multi-parameter estimation, such as spectroscopy, electromagnetic field sensing and gravitational wave field detection, it is worthy to apply chaotic dynamics into these existing systems.

\acknowledgements{This work is supported by the Key-Area Research and Development Program of GuangDong Province under
Grants No. 2019B030330001, the NSFC (Grant No. 11874434, No. 11574405, and No. 11704420), and the Science and Technology Program of Guangzhou (China) under Grants No. 201904020024.}
\appendix

\section{State distribution in Bloch sphere} \label{appendixA}
\begin{figure}[htp]
\begin{center}
\includegraphics[width=0.45\textwidth]{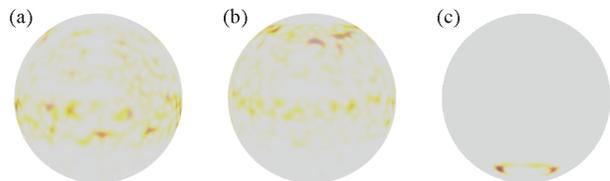}
\end{center}
\caption{A distribution comparison of different final states, occurring for different initial states labeled in Fig.~\ref{fig:2}~(d). Different from the widely spreading in generalized Bloch sphere for chaotic sea in (a) and edge state in (b), the regular dynamics is restricted in a localized region for regular island in (c).
}
\label{fig:Bloch}
\end{figure}

The distribution in Bloch sphere provides a natural way to display the state difference.
Based on three different initial parameters labeled in Fig.~\ref{fig:2} (d),
after a long time-evolution up to $t=2^{15}T$, the obtained final state as well as its density matrix
are described by $|\psi_f\rangle$ and $\rho=|\psi_f\rangle\langle\psi_f|$, respectively.
One can project the final state $|\psi_f\rangle$ into the $Q$ representation
\begin{equation}
Q(\theta, \phi)=\frac{2 J+1}{4 \pi}\langle\theta, \phi|\rho| \theta, \phi\rangle ,
\end{equation}
then vividly takes on the generalized Bloch sphere, see Fig.~\ref{fig:Bloch}.
The initial states in chaotic region and edge region are scrambled into the entire system and lost its local information during the process of quantum chaos, as shown in Figs.~\ref{fig:Bloch} (a) and (b), respectively.
Comparing to chaotic regime, the distribution of final state launched from a
regular island is restricted within this stability island, see Fig.~\ref{fig:Bloch} (c).
A discrepancy is evident for both the regular and chaotic initial SCS.

\section{Size scaling analysis} \label{appendixB}

\begin{figure}[htp]
\begin{center}
\includegraphics[width=0.45\textwidth]{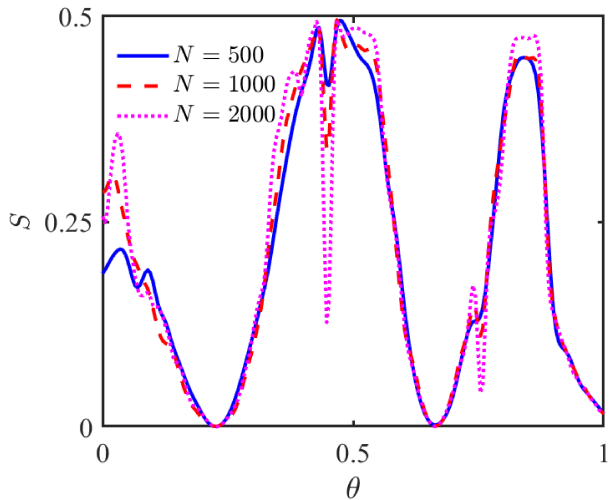}
\end{center}
\caption{The size scaling analysis about linear entropy versus  polar angle is performed with the large accessible system size $N=500$ (solid line), $1000$ (dashed line) and $2000$ (dotted line).}
\label{fig:size}
\end{figure}

The phase spaces in quantum setting focus on the finite-$N$ systems, as shown in Figs.~\ref{fig:2} (b)-(d).
To investigate the size scaling analysis more intuitively, we extract all points on the line $\phi=\pi$ in Fig.~\ref{fig:2} (b) as initial states for numerical simulation, and calculate the linear entropy as function of $\theta$ in Fig.~\ref{fig:size}.
It can be clearly observed that the minimum of linear entropy corresponds to the classic fixed point in the regular region, while its maximum corresponds to the region with the strongest chaos in the classical phase space.
The different system sizes are denoted as different symbols.
A transition from regions of high linear entropy to low one becomes more stark with increasing system size, making the discrepancy between chaotic and regular region more abruptly.
%

\bibliographystyle{apsrev4-1}

\end{document}